\documentclass[%
 reprint,
 amsmath,amssymb,
 aps,
]{revtex4-2}

\usepackage{graphicx}
\usepackage{dcolumn}
\usepackage{bm}

\begin{document}

\preprint{APS/123-QED}

\title{Sign freedom of non-Abelian topological charges in phononic and photonic topological insulators}

\author{Haedong Park}
 \affiliation{
 School of Physics and Astronomy, Cardiff University, Cardiff CF24 3AA, United Kingdom
 }
\author{Sang Soon Oh}%
 \email{ohs2@cardiff.ac.uk}
\affiliation{
 School of Physics and Astronomy, Cardiff University, Cardiff CF24 3AA, United Kingdom
}%
\date{\today}

\begin{abstract}
The topological nature of nodal lines in a three-band system can be described by non-Abelian topological charges called quaternion numbers.
Due to the gauge freedom of the eigenstates, the sign of the non-Abelian topological charges can be flipped by performing the gauge transformation, i.e., choosing a different basis of eigenstates. 
However, the sign flipping has not been explicitly shown in realistic systems such as phononic and photonic topological insulators.  
Here, we elaborate the sign freedom by visualizing the numerically calculated topological charges in phononic and photonic topological insulators.
In doing so, we employ a common reference point method for multiple nodal lines to confirm that the sign flipping does not cause any issue in building the quaternion group.

\end{abstract}

\maketitle

\section{\label{sec:Introduction}Introduction}
Multiple nodal lines in the momentum space can be interpreted by non-Abelian band topology \cite{Wu_2019_Science}. In particular, topological charges (invariants) of nodal lines in a three-band system can be expressed by the quaternion group ${\mathbb Q} = \left\{ \pm \boldsymbol i, \pm \boldsymbol j, \pm \boldsymbol k, \pm 1 \right\}$ \cite{Wu_2019_Science}. 
Recently, a lot of studies on the nodal lines \cite{Lenggenhager_2021_PRB,Bouhon_2020_NatPhys,Xia_2019_PRL,Ahn_2018_PRL,Kim_2020_LSA}
have been reported, including nodal rings \cite{Deng_2019_NatComm,Gao_2018_NatComm}, nodal chains
\cite{Bzdusek_2016_Nature,Yan_2018_NatPhys,Chang_2017_PRL,Yan_2017_PRB,Belopolski_2019_Science},
nodal links \cite{Chang_2017_PRL,Yan_2017_PRB,Belopolski_2019_Science,YangErchan_2020_PRL_SZhang,He_2020_PRA,Xie_2019_PRB,Lee_2020_NatComm,YangZhesen_2020_PRL_ZonesPolynomial}, and nodal knots
\cite{Lee_2020_NatComm,YangZhesen_2020_PRL_ZonesPolynomial,Bi_2017_PRB}. 
Theoretical and experimental efforts have been made to demonstrate the quaternion topological charges in such nodal line systems \cite{Park_2021a_ACSPhotonics,YangErchan_2020_PRL_SZhang,Wang_2021_LSA,Yang_2021_NatComm,Guo_2021_Nature}. Hence, the field of research on the non-Abelian topological nodal lines is rapidly expanding along with the study on Dirac
\cite{Slobozhanyuk_2017_NatPhoton,Lu_2016_NatPhys,Jin_2017_PRL,Liu_2020_NatComm,Abbaszadeh_2017_PRL,Brendel_2017_PNAS,Wang_2015_PRL,Wen_2019_NatPhys}
and Weyl points
\cite{Lu_2013_NatPhoton,Lu_2015_Science,Park_2020_ACSPhotonics,He_2018_Nature,Peri_20190_NatPhys,He_2020_NatComm,Jia_2019_Science,Yang_2018_Science,Yang_2019_NatComm,Soluyanov_2015_Nature}.

Although the gauge-dependent property of the non-Abelian charge has been already discussed \cite{Wu_2019_Science,Bouhon_2019_PRB,Bouhon_2020_PRB,Tiwari_2020_PRB,Guo_2021_Nature,Jiang_2021_NatPhys,Peng_2021_arXiv}, such a property has not been explained by visualizing the eigenstates. In any eigenvalue problems, if $\mathbf u$ is an eigenstate of a system, $e^{i \theta} {\mathbf u}$ ($\theta = \left(0, 2\pi \right]$) is also an eigenstate of the system, and these two are regarded as the equivalent states (gauge freedom). Even if we assume only a real number $e^{i \theta}$, i.e., $\theta=0$ or $\pi$, the eigenstates of the system can be expressed either by $\mathbf u$ or $-\mathbf u$. This gauge freedom may generate any signs of the topological invariants in the nodal line system. Notably, this property does not contradict the argument of the non-Abelian band topology in a three-band system and it is important to fix the basis of eigenstates as pointed out in Ref.~\cite{Wu_2019_Science}.  

In this paper, we elucidate the gauge-dependent sign freedom of the non-Abelian topological charges by visualizing the topological charges in phononic and photonic topological insulators. First, we exemplify such a property with phononic and photonic systems. Then, the reference point method which can be thought as fixing the basis of Hilbert space \cite{Wu_2019_Science,Tiwari_2020_PRB} is adopted to our topological insulators. Finally, we discuss the consistency of the topological charges’ non-Abelian relations and the considerations of using this method in phononic and photonic topological insulators regarding the non-Abelian topological charges.

\section{\label{sec:Visualization}Gauge-dependent property of topological charges}
Obtaining the quaternion charge of a nodal line starts by calculating eigenstates along the closed loop that encloses the nodal line.
We denote the starting point of the closed loop’s winding as ${\mathbf k}_0$.
Once the signs of the eigenstates at ${\mathbf k}_0$ are determined, their signs at subsequent points $\mathbf k$ on the loop are assigned.
Finally, the topological charges are calculated by analyzing the rotation behaviors of the eigenstates.
Due to the sign freedom of the eigenstates at ${\mathbf k}_0$, the topological charge also has the sign freedom \cite{Wu_2019_Science,Bouhon_2019_PRB,Bouhon_2020_PRB,Tiwari_2020_PRB,Guo_2021_Nature,Jiang_2021_NatPhys,Peng_2021_arXiv}.
In this section, we discuss the examples that show the sign freedom of the quaternion charges originating from the sign freedom of eigenstates at ${\mathbf k}_0$.

\begin{figure*}
    \includegraphics{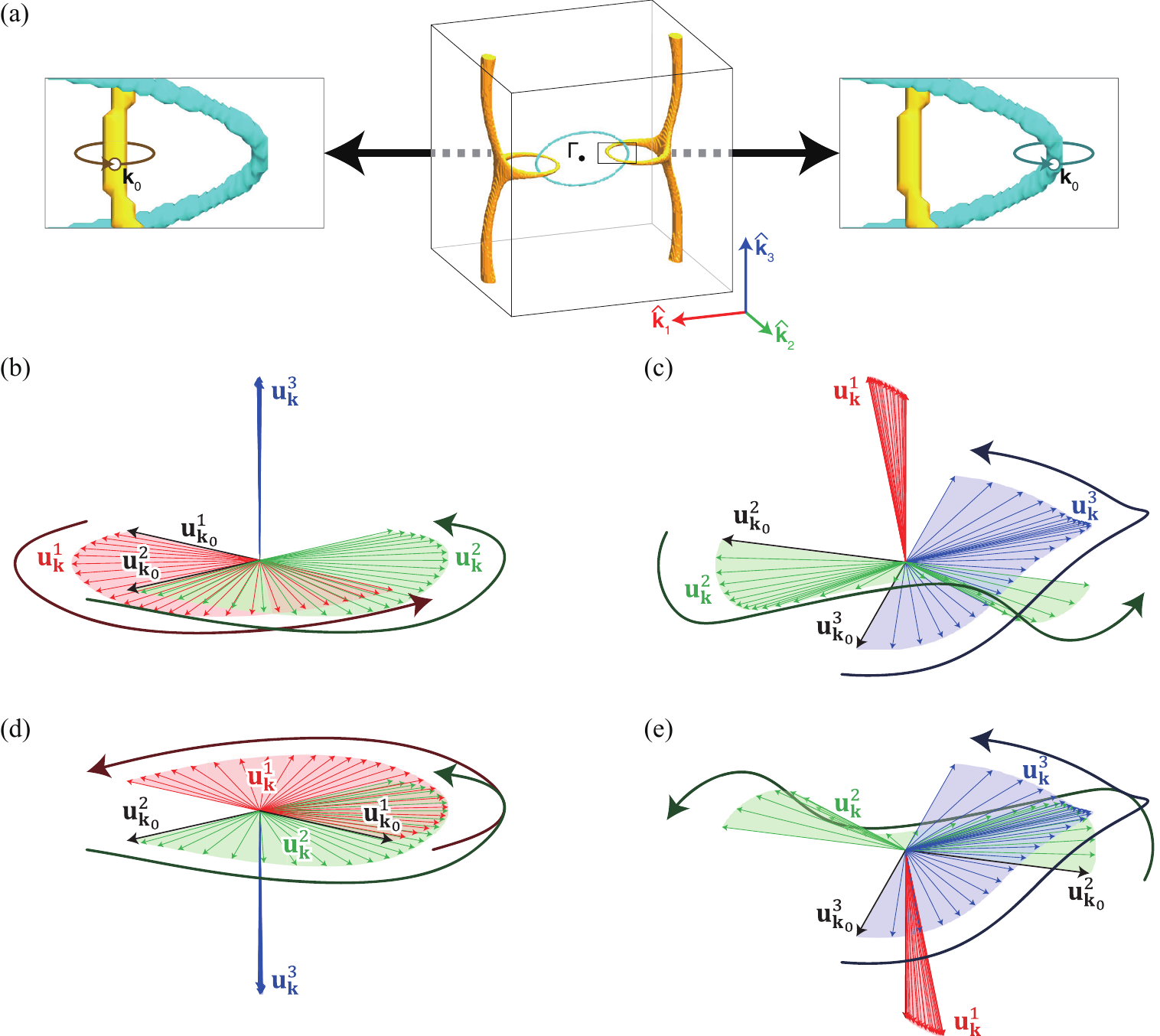}
    \caption{
        \label{fig:Example_3x3}
            An example of $3\times3$ Hamiltonian about elastic waves that exhibits either positive or negative topological charges for a single closed loop.
            (a) Nodal link by the elastic waves (middle) and the closed loops that encircle the orange (left) and cyan (right) nodal lines.
            The orange and cyan nodal rings are formed by the first and second bands and the second and third bands, respectively.
            The white open circles on the closed loops mean the points ${\mathbf k}_0$, and the arrows indicate the winding directions.
            (b),(c) Eigenvectors ${\mathbf u}_{\mathbf k}^n$ obtained along the closed loop in the left and right figures in (a), respectively.
            (d) Eigenvectors replotted after flipping the signs of ${\mathbf u}_{{\mathbf k}_0}^1$ and ${\mathbf u}_{{\mathbf k}_0}^3$ in (b).
            (e) Eigenvectors replotted after flipping the signs of ${\mathbf u}_{{\mathbf k}_0}^1$ and ${\mathbf u}_{{\mathbf k}_0}^2$ in (c).
    }
\end{figure*}
\begin{figure*}
    \includegraphics{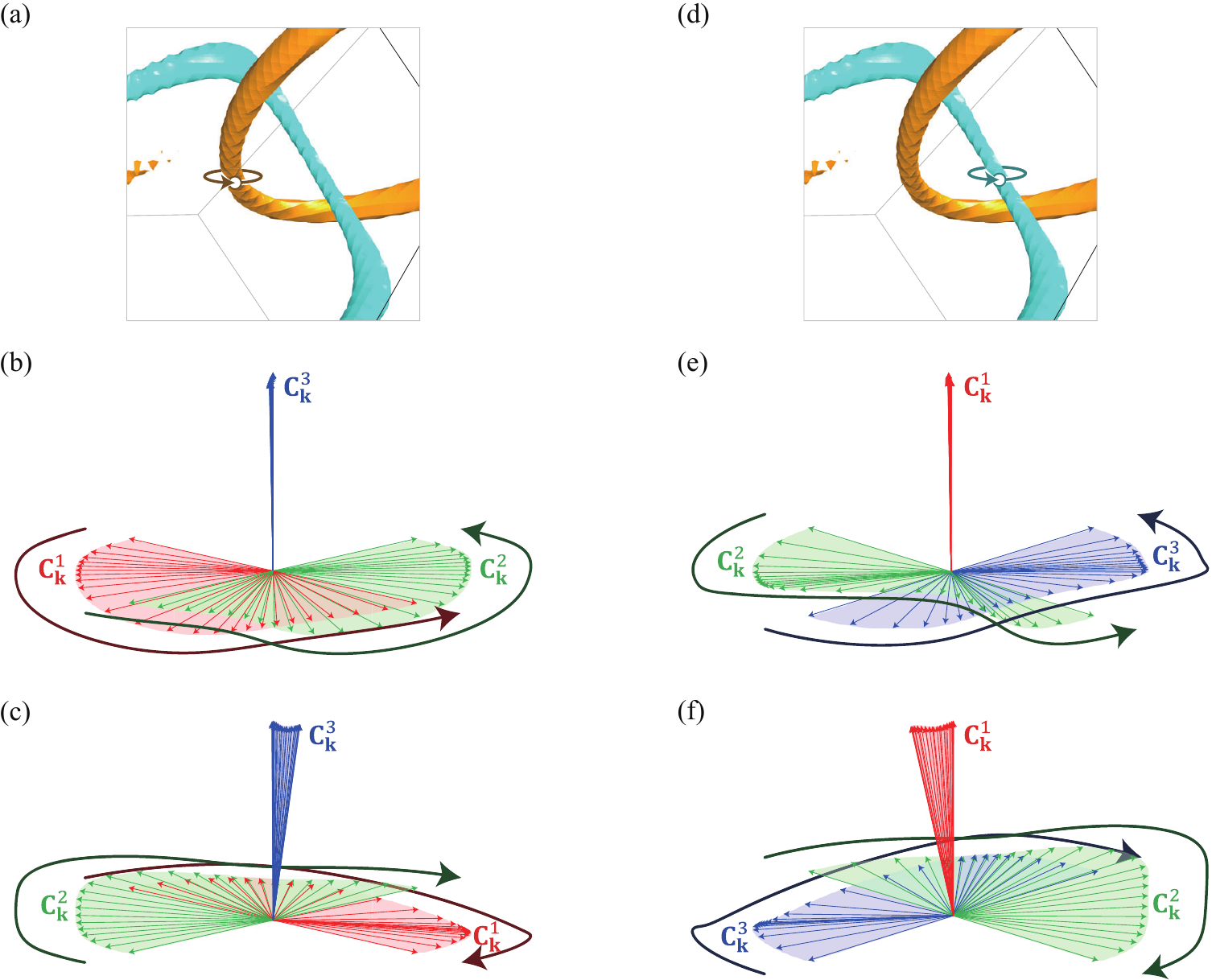}
    \caption{
        \label{fig:Example_Correlation}
        Comparison of the topological charge of the nodal lines in a photonic system \cite{Park_2021a_ACSPhotonics} when the correlations’ signs are manipulated.
        (a),(d) Closed loops that enclose the orange and cyan nodal rings, respectively \cite{Park_2021a_ACSPhotonics}.
        The orange and cyan nodal rings are formed by the third and fourth bands and the fourth and fifth bands, respectively.
        Adapted from Park et al, \emph{ACS Photonics} 2021 \cite{Park_2021a_ACSPhotonics}. 
        (b),(c) Correlations ${\mathbf C}_{\mathbf k}^n$ without and with the sign-flipping of $\left| \psi_{\mathbf k}^1 \right\rangle$ and $\left| \psi_{\mathbf k}^3 \right\rangle$, respectively.
        (e),(f) Correlations ${\mathbf C}_{\mathbf k}^n$ without and with the sign-flipping of $\left| \psi_{\mathbf k}^1 \right\rangle$ and $\left| \psi_{\mathbf k}^2 \right\rangle$, respectively.
    }
\end{figure*}
\begin{figure}
    \includegraphics{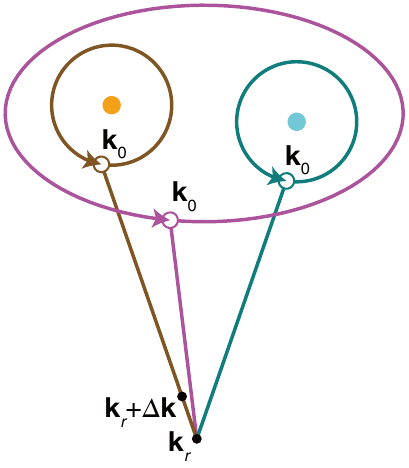}
    \caption{
        \label{fig:RefPoint}
        Reference point for the gauge fixing. ${\mathbf k}_r$ is the common reference point. The lines that connect ${\mathbf k}_r$ and ${\mathbf k}_0$ are also shown. Nodal lines are marked as orange and cyan points.
    }
\end{figure}
\begin{figure*}
    \includegraphics{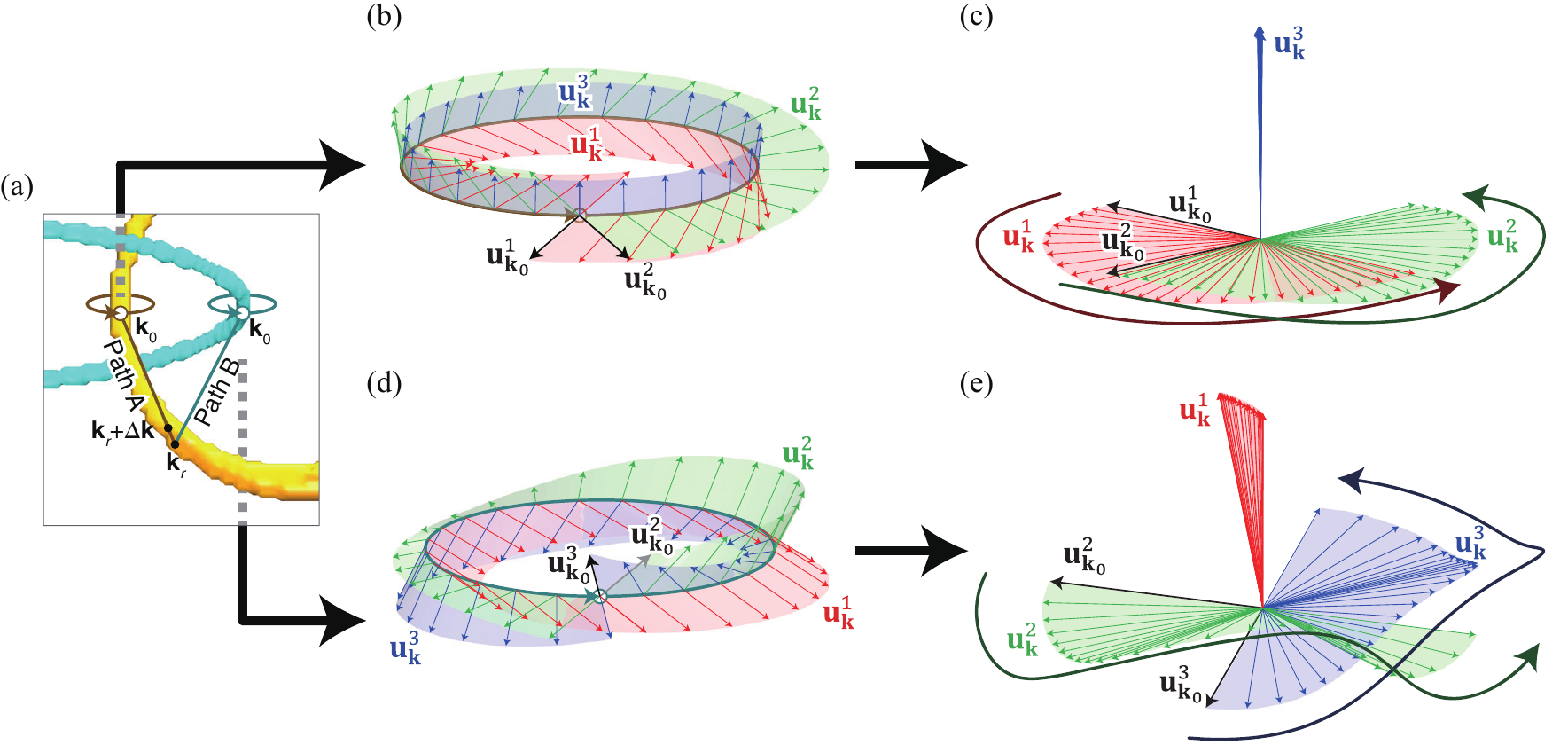}
    \caption{
        \label{fig:k_i_AfterSignConvention}
        Evaluation of the topological charges ${\boldsymbol k}$ and ${\boldsymbol i}$.
        (a) Two closed loops encircling each nodal line and the paths that connect ${\mathbf k}_0$ of each closed loop and ${\mathbf k}_r$.
        (b),(d) Eigenstates along the closed loops plotted with respect to an arbitrary orthonormal coordinate system.
        (c),(e) Eigenstates collections at the origin. They indicate their topological charges as $+{\boldsymbol k}$ and $+{\boldsymbol i}$, respectively.
    }
\end{figure*}

\subsection{\label{sec:example_3x3}$3\times3$ Hamiltonian}

We consider phononic waves in an orthotropic elastic material \cite{Beer_2012_Book} whose constitutive equation is given by Hooke's law ${\boldsymbol \varepsilon} = {\cal C} \cdot {\mathbf s}$. Here, ${\boldsymbol \varepsilon} = \left\{ \varepsilon_{ij} \right\}$ and ${\mathbf s} = \left\{ s_{ij} \right\}$ ($i,j=1,2,3$) are the Cauchy strain and stress tensors, respectively.
$\cal C$ is a compliance matrix, which is a function of Young's moduli, shear moduli and Poisson's ratios: ${\cal C} = {\cal C} \left( E_i , G_{ij} , \nu_{ij} \right)$. Its inverse is the stiffness matrix ${\cal S}$, i.e., ${\cal C}^{-1} = {\cal S}$, thereby the consitutive equation becomes ${\mathbf s} = {\cal S} \cdot {\boldsymbol \varepsilon}$ which is analogous to the one-dimensional Hooke's law $F=-kx$.

By using the density $\rho$, Young's moduli $E_i$, shear moduli $G_{ij}$, Poisson’s ratios $\nu_{ij}$, and restoring forces $f_i$ ($i,j=1,2,3$) listed in Table \ref{tab:ParameterData} in Appendix~\ref{OrthotropicDerivation}, the eigenvalue problem for this elastic wave system is expressed as
\begin{widetext}
    \begin{equation}
        \left[
            \begin{array}{ccc}
                k_1^2 M_{11}+k_2^2 G_{12}+k_3^2 G_{31}+f_1 & k_1 k_2\left(\lambda_{12}+G_{12}\right) & k_1 k_3\left(\lambda_{31}+G_{31}\right) \\
                k_1 k_2\left(\lambda_{12}+G_{12}\right) & k_1^2 G_{12}+k_2^2 M_{22}+k_3^2 G_{23}+f_2 & k_2 k_3\left(\lambda_{23}+G_{23}\right) \\
                k_1 k_3\left(\lambda_{31}+G_{31}\right) & k_2 k_3\left(\lambda_{23}+G_{23}\right) & k_1^2 G_{31}+k_2^2 G_{23}+k_3^2 M_{33}+f_3
            \end{array}
        \right]
        {\mathbf u}_{\mathbf k}^n = \rho \left( \omega^n \right)^2 {\mathbf u}_{\mathbf k}^n ,
    \label{orthogropic3x3}
    \end{equation}
\end{widetext}
where ${\mathbf u}_{\mathbf k}^n$ ($n=1,2,3$) is the three-component displacement eigenvector satisfying the orthonormal condition ${\mathbf u}_{\mathbf k}^m \cdot {\mathbf u}_{\mathbf k}^n =\delta ^{m n} $, and $\omega^n$ is the angular frequency as an eigenfrequency [the detailed derivation of Eq.~(\ref{orthogropic3x3}) is written in Appendix~\ref{OrthotropicDerivation}].
Here, the superscript $n$ is the band number, and $\mathbf k$ is a point in the momentum space.

For the nodal link shown in Fig.~\ref{fig:Example_3x3}(a) in this phononic system, we choose two closed loops [the arrowed loops in Fig.~\ref{fig:Example_3x3}(a)] that enclose each nodal ring [the orange and cyan colored shapes Fig.~\ref{fig:Example_3x3}(a), respectively].
We calculate the orthonormal eigenvectors ${\mathbf u}_{\mathbf k}^n$ along the closed loops \cite{Park_2021a_ACSPhotonics}.
We gather all the eigenstates at the origin, and their behaviors are shown in Fig.~\ref{fig:Example_3x3}(b), (c), respectively.
We may regard the topological charge for Fig.~\ref{fig:Example_3x3}(b) as the quaternion number $\boldsymbol k$ because ${\mathbf u}_{\mathbf k}^1$ and ${\mathbf u}_{\mathbf k}^2$ rotate by $+\pi$ around fixed ${\mathbf u}_{\mathbf k}^3$ \cite{Wu_2019_Science,Park_2021a_ACSPhotonics,YangErchan_2020_PRL_SZhang}.
Likewise, ${\mathbf u}_{\mathbf k}^2$ and ${\mathbf u}_{\mathbf k}^3$ rotate by $+\pi$ around fixed ${\mathbf u}_{\mathbf k}^1$ in Fig.~\ref{fig:Example_3x3}(c), and thereby its topological charge is the quaternion number $\boldsymbol i$ \cite{Wu_2019_Science,Park_2021a_ACSPhotonics,YangErchan_2020_PRL_SZhang}.

Meanwhile, due to the sign freedom of the eigenstates, we can choose different eigenstates at ${\mathbf k}_0$.
For the orange loop in Fig.~\ref{fig:Example_3x3}(a), we flip the signs of ${\mathbf u}_{{\mathbf k}_0}^1$ and ${\mathbf u}_{{\mathbf k}_0}^3$ while keeping ${\mathbf u}_{{\mathbf k}_0}^2$ fixed.
Then, ${\mathbf u}_{\mathbf k}^n$ at the following points ${\mathbf k}$ on the loop is determined by these ${\mathbf u}_{{\mathbf k}_0}^n$.
The resulting relation between the eigenstates ${\mathbf u}_{\mathbf k}^n$ exhibits $-\pi$-rotations of ${\mathbf u}_{\mathbf k}^1$ and ${\mathbf u}_{\mathbf k}^2$ around ${\mathbf u}_{\mathbf k}^3$ [see Fig.~\ref{fig:Example_3x3}(d)].
Thus, the topological charge for this situation is considered as $-\boldsymbol k$, compared to Fig.~\ref{fig:Example_3x3}(b).
Likewise, we apply the sign-flipping of ${\mathbf u}_{{\mathbf k}_0}^1$ and ${\mathbf u}_{{\mathbf k}_0}^2$ with fixing ${\mathbf u}_{{\mathbf k}_0}^3$ to the cyan loop in Fig.~\ref{fig:Example_3x3}(a).
It generates the $-\pi$-rotations of ${\mathbf u}_{\mathbf k}^2$ and ${\mathbf u}_{\mathbf k}^3$ around ${\mathbf u}_{\mathbf k}^1$ [see Fig.~\ref{fig:Example_3x3}(e)] making its topological charge $-\boldsymbol i$ in contrast to Fig.~\ref{fig:Example_3x3}(c).
Therefore, the topological charges of the left and right figures of Fig.~\ref{fig:Example_3x3}(a) can be any of the four sets of quaternion numbers
$\left[ {\boldsymbol k}, {\boldsymbol i} \right]$,
$\left[ -{\boldsymbol k}, {\boldsymbol i} \right]$,
$\left[ {\boldsymbol k}, - {\boldsymbol i} \right]$, or
$\left[- {\boldsymbol k}, - {\boldsymbol i} \right]$.

\subsection{\label{sec:example_Correlation}Correlation vectors in photonic system}

Examples of the gauge-dependent non-Abelian charge can be also visualized by the correlation vectors \cite{Park_2021a_ACSPhotonics}.
For the closed loop in Fig.~\ref{fig:Example_Correlation}(a), we compute the orthonormal eigenstates $\left| \psi_{\mathbf k}^n \right\rangle$ ($n=1,2,3$ for the third, fourth, and fifth bands, respectively).
Then, we calculate the correlations ${\mathbf C}_{\mathbf k}^n$ defined by the following equation \cite{Park_2021a_ACSPhotonics}:
\begin{equation}
    {\mathbf C}_{\mathbf k}^n
    =
    \left[
        \left\langle
        \psi_{{\mathbf k}_0}^1
        |
        \psi_{\mathbf k}^n
        \right\rangle
        ,
        \left\langle
        \psi_{{\mathbf k}_0}^2
        |
        \psi_{\mathbf k}^n
        \right\rangle
        ,
        \left\langle
        \psi_{{\mathbf k}_0}^3
        |
        \psi_{\mathbf k}^n
        \right\rangle
    \right]
\end{equation}
All the correlations ${\mathbf C}_{\mathbf k}^n$ on the closed loop are collected at the origin, as shown in Fig.~\ref{fig:Example_Correlation}(b).
The result means its topological charge is $-i \sigma_3 $.
Now, we adjust the signs of $\left| \psi_{{\mathbf k}_0}^1 \right\rangle$ and $\left| \psi_{{\mathbf k}_0}^3 \right\rangle$ with fixing the sign of $\left| \psi_{{\mathbf k}_0}^2 \right\rangle$.
The signs of $\left| \psi_{\mathbf k}^n \right\rangle$ at the remaining points $\mathbf k$ on the closed loop are assigned by the $\left| \psi_{{\mathbf k}_0}^n \right\rangle$.
The resulting topological charge in Fig.~\ref{fig:Example_Correlation}(c) is $+i \sigma_3 $.

Likewise, for the closed loop in Fig.~\ref{fig:Example_Correlation}(d), we consider two sets of $\left| \psi_{{\mathbf k}_0}^n \right\rangle$.
The sign of $\left| \psi_{{\mathbf k}_0}^n \right\rangle$ in the first set is not adjusted.
For the second set, we flip the signs of $\left| \psi_{{\mathbf k}_0}^1 \right\rangle$ and $\left| \psi_{{\mathbf k}_0}^2 \right\rangle$ with fixing the gauge of $\left| \psi_{{\mathbf k}_0}^3 \right\rangle$.
Thus, the topological charges of Fig.~\ref{fig:Example_Correlation}(e), (f) are $-i \sigma_1 $ and $+i \sigma_1 $, respectively.

\begin{figure*}
    \includegraphics{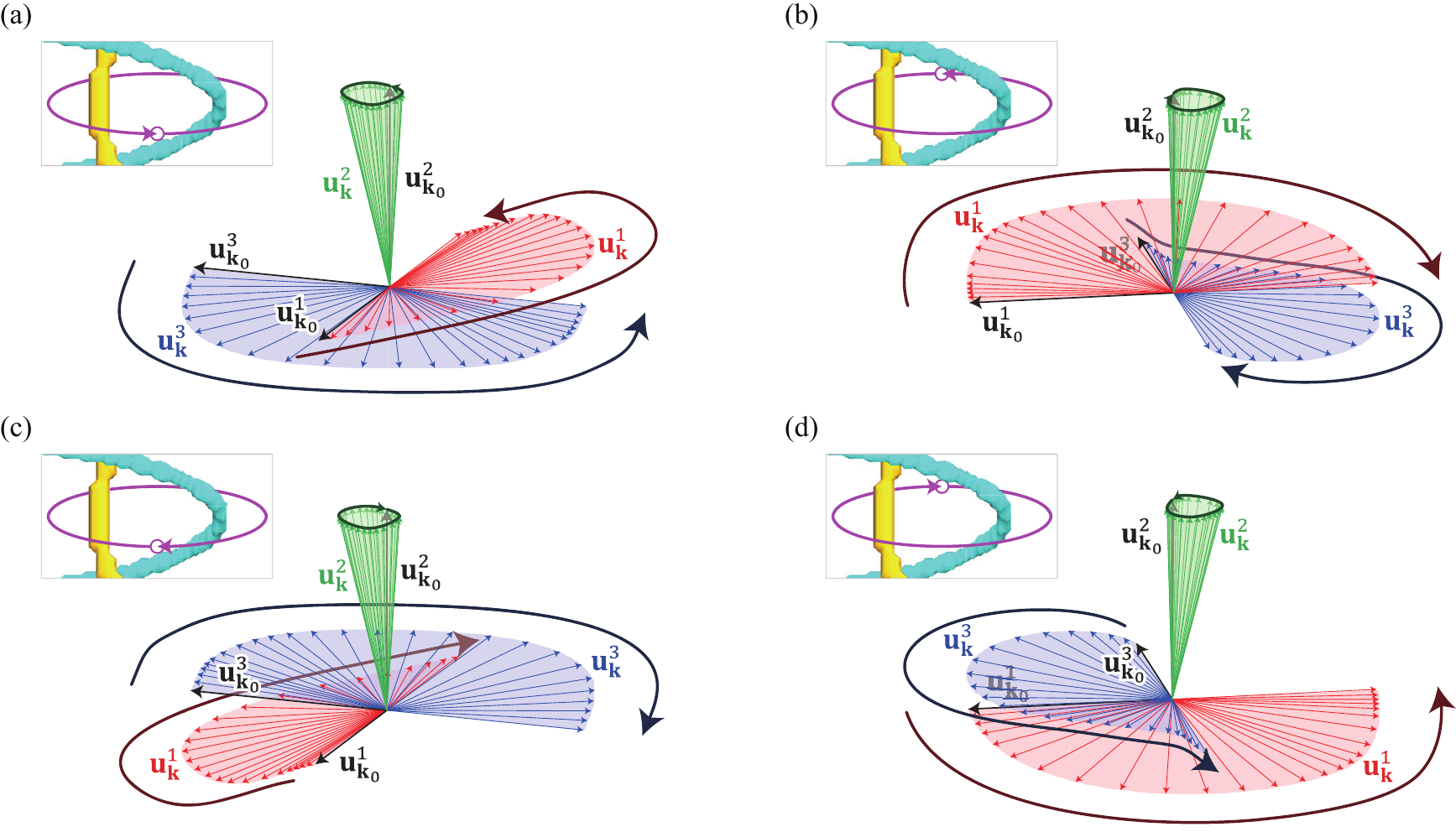}
    \caption{
        \label{fig:j_AfterSignConvention}
        Topological charge ${\boldsymbol j}$ by the closed loop encircling both orange and cyan nodal lines. Each panel corresponds to ${\boldsymbol k} {\boldsymbol i} = {\boldsymbol j}$, ${\boldsymbol i} {\boldsymbol k} = -{\boldsymbol j}$, $\left( -{\boldsymbol i} \right) \left( -{\boldsymbol k}\right) = -{\boldsymbol j}$, and $\left( -{\boldsymbol k} \right) \left( -{\boldsymbol i}\right) = {\boldsymbol j}$, respectively.
    }
\end{figure*}
\begin{figure}
    \includegraphics{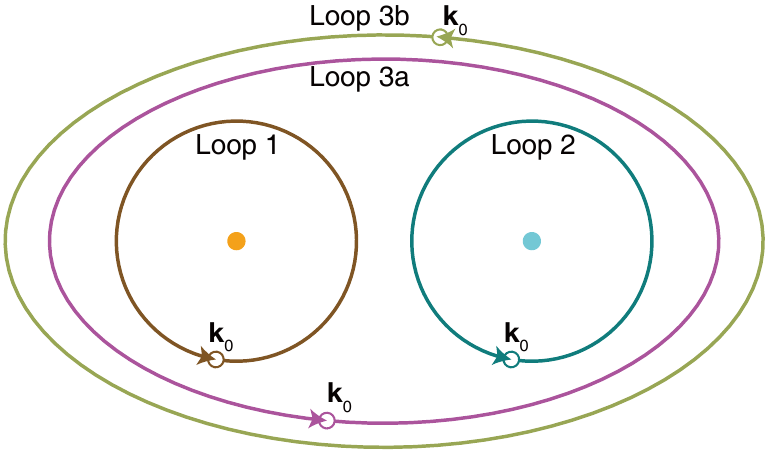}
    \caption{
        \label{fig:N1N2N3}
        Closed loops composition about ${\mathbf N}_{3{\mathrm a}} = {\mathbf N}_1 {\mathbf N}_2$ and ${\mathbf N}_{3{\mathrm b}} = {\mathbf N}_2 {\mathbf N}_1$.
    }
\end{figure}
\begin{figure*}
    \includegraphics{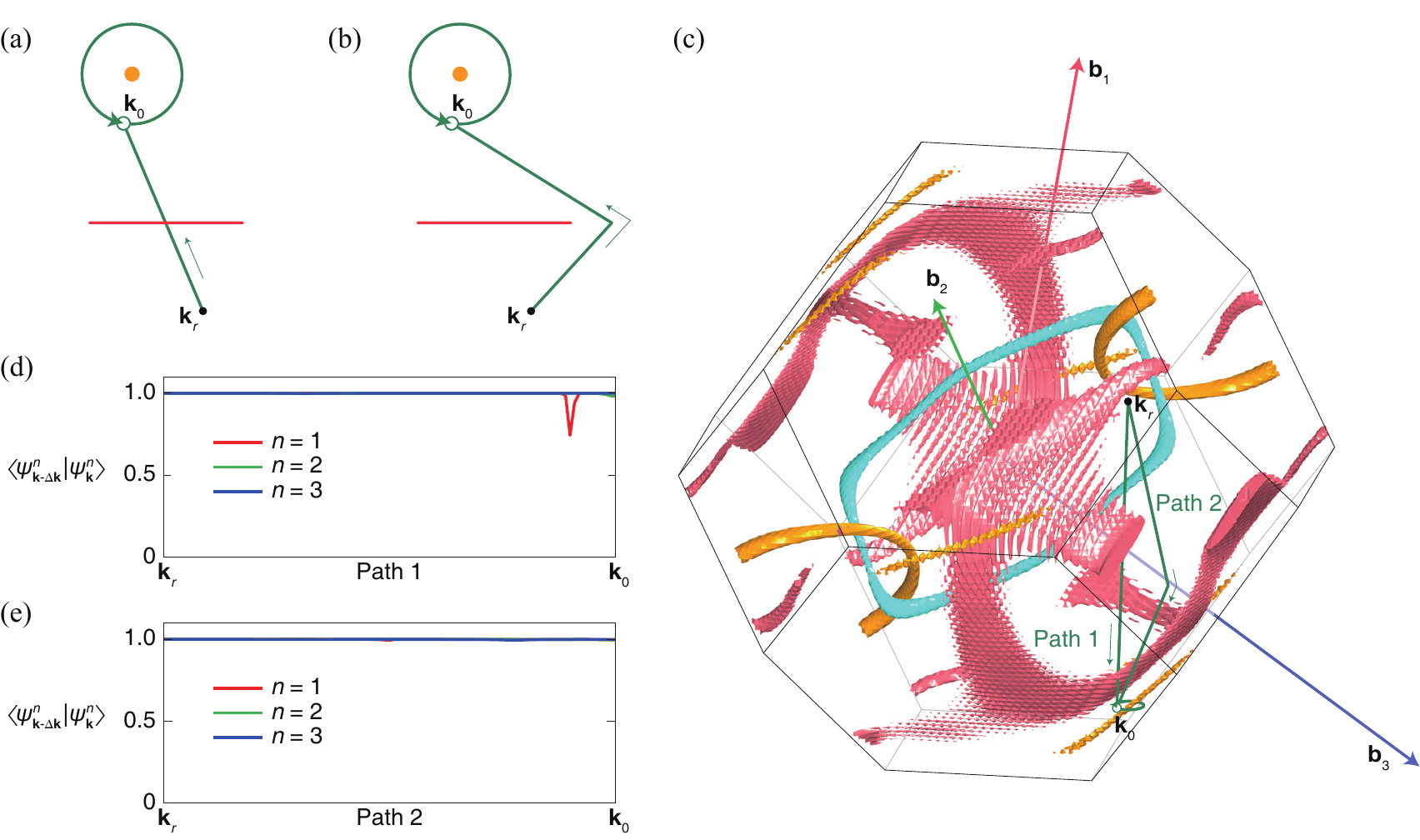}
    \caption{
        \label{fig:PathSetting}
        Consideration for using the reference point.
        (a),(b) Two examples of ${\mathbf k}_r - {\mathbf k}_0$ paths going through or avoiding the degeneracies, respectively.
        Red lines mean degeneracies.
        (c) Nodal link and another set of degeneracies by the double diamond structure \cite{Park_2021a_ACSPhotonics}.
        The nodal link is formed by the third, fourth, and fifth bands ($n = 1$, $2$, and $3$, respectively), and the pink-colored degeneracies are formed by the second and third bands ($n = 0$ and $1$, respectively).
        Two paths are also marked. Path 1 intersects the pink degeneracies near ${\mathbf k}_0$ while Path 2 does not.
        (d),(e) $\left\langle \psi_{{\mathbf k}-\Delta {\mathbf k}}^n \mid \psi_{\mathbf k}^n \right\rangle$ along the Paths 1 and 2, respectively.
    }
\end{figure*}
\section{\label{sec:CommonReference}Application of common reference point method}

To fix the signs of the topological charges of our topological insulators, we fix the gauge of the eigenstates using a reference point \cite{Wu_2019_Science,Tiwari_2020_PRB}.
We denote the common reference point as ${\mathbf k}_r$, and a line connects ${\mathbf k}_r$ and ${\mathbf k}_0$ of each loop, as illustrated in Fig.~\ref{fig:RefPoint}.
There can be several ${\mathbf k}_0$ for several closed loops while we consider only one ${\mathbf k}_r$.
For example, when the topological charge on one closed loop in Fig.~\ref{fig:RefPoint} is investigated, $\left| \psi_{{\mathbf k}_r}^n \right\rangle$ is calculated first.
For all the remaining points on the line, $\left| \psi_{{\mathbf k}_r +\Delta {\mathbf k}}^n \right\rangle$ is also calculated.
Finally, when ${\mathbf k} +\Delta {\mathbf k}$ reaches ${\mathbf k}_0$, $\left| \psi_{{\mathbf k}_0}^n \right\rangle$ is determined. 
Then, the topological charge is calculated by the eigenstates only on the loop.

Instead of using the reference point ${\mathbf k}_r$, choosing the loops whose starting points ${\mathbf k}_0$ coincide can be an alternative.
However, a smaller loop generates more accurate topological charge.
Thus, this alternative can be used in limited situations only, for example the nodal lines that are sufficiently close to each other.

For the phononic topological insulator, we set a common reference point ${\mathbf k}_r = \left[ -1,1,0.5 \right]$ as shown in Fig.~\ref{fig:k_i_AfterSignConvention}(a), then we calculated the eigenstates of Eq.~(\ref{orthogropic3x3}) along each path.
The eigenstates on the closed loop of Path A are shown in Fig.~\ref{fig:k_i_AfterSignConvention}(b), and collecting them at the origin results in $+{\boldsymbol k}$ [as shown in Fig.~\ref{fig:k_i_AfterSignConvention}(c), identical to Fig.~\ref{fig:Example_3x3}(b)].
Likewise, the eigenstates along Path B are plotted in Fig.~\ref{fig:k_i_AfterSignConvention}(d) and (e) [identical to Fig.~\ref{fig:Example_3x3}(c)] resulting in topological charge $+{\boldsymbol i}$. The eigenstates for Path A and B should be the same at the common reference point ${\mathbf k}_r$.

The topological charge ${\boldsymbol j} = {\boldsymbol k} {\boldsymbol i}$ can be described by composition of the loops in Path A and B, which are the closed loops encircling both orange and cyan nodal lines.
Here, the circling sequence is related to the order of ${\boldsymbol k}$ and ${\boldsymbol i}$ in the multiplication.
If the first and second half of closed loop scan the cyan (${\boldsymbol i}$) and orange (${\boldsymbol k}$) nodal lines, respectively, the relation is written as ${\boldsymbol k} {\boldsymbol i}$.
The signs of ${\boldsymbol k}$ and ${\boldsymbol i}$ are concerned with winding direction of the closed loop.
Then, we can consider four relations ${\boldsymbol k} {\boldsymbol i}$, ${\boldsymbol i} {\boldsymbol k}$, $\left( -{\boldsymbol i} \right) \left( -{\boldsymbol k}\right)$, and $\left( -{\boldsymbol k} \right) \left( -{\boldsymbol i}\right)$, and their results are thus ${\boldsymbol j}$, $-{\boldsymbol j}$, $-{\boldsymbol j}$, and ${\boldsymbol j}$, as shown in each panel of Fig.~\ref{fig:j_AfterSignConvention}, respectively.
The plots in Fig.~\ref{fig:j_AfterSignConvention} are the calibrated results by Ref.~\cite{Park_2021a_ACSPhotonics}.
The calibration angle was set as $0.8 \theta_0 \left( {\mathbf k} \right)$ where $\theta_0 \left( {\mathbf k} \right)$ is the angle between ${\mathbf u}_{\mathbf k}^2$ and ${\mathbf u}_{{\mathbf k}_0}^2$.

Note that the topological charges ${\boldsymbol k}$, ${\boldsymbol i}$, and ${\pm \boldsymbol j}$ of the double diamond photonic crystal in Ref.~\cite{Park_2021a_ACSPhotonics} were calculated using the reference point. More details will be explained in Section~\ref{sec:PathSetting}.

\section{\label{sec:Consistency_RefPoint}Consistency of topological charges’ signs by the reference point method}

In the following, we will elaborate that the reference point method can eliminate the chance that the non-Abelian charges are misinterpreted as Abelian.
For example, let us suppose that two closed loops are in different locations [marked as Loop 1 and 2 in Fig.~\ref{fig:N1N2N3}, respectively].
If we denote topological charges by these loops as ${\mathbf N}_1$ and ${\mathbf N}_2$, respectively, and if we consider an additional loop [marked as Loop 3a in Fig.~\ref{fig:N1N2N3}] circling along Loop 2 and 1 in sequence \cite{Park_2021a_ACSPhotonics,YangErchan_2020_PRL_SZhang}, the Loop 3a’s topological charge ${\mathbf N}_{3{\mathrm a}}$ is calculated by the composition of ${\mathbf N}_1$ and ${\mathbf N}_2$, i.e., ${\mathbf N}_{3{\mathrm a}} = {\mathbf N}_1 {\mathbf N}_2$.
We also suppose that Loop 3b scans around Loop 1 first then circles Loop 2, as shown in Fig.~\ref{fig:N1N2N3}. Its topological charge is expressed as ${\mathbf N}_{3{\mathrm b}} = {\mathbf N}_2 {\mathbf N}_1$.
If ${\mathbf N}_{3{\mathrm a}}$ and ${\mathbf N}_{3{\mathrm b}}$ satisfy ${\mathbf N}_{3{\mathrm a}} = - {\mathbf N}_{3{\mathrm b}}$, this leads to ${\mathbf N}_1 {\mathbf N}_2 = - {\mathbf N}_2 {\mathbf N}_1$. This relation can be called `non-Abelian'.
To write this relation, the signs of each charge should be first determined by such kind of sign convention.
If there is not any sign convention, and if the topological charge of Loop 3b can be either plus or minus, it means that we can assume the sign-flipping of ${\mathbf N}_{3{\mathrm b}}$ with fixing the other charges' signs. Then, the above relation ${\mathbf N}_{3{\mathrm a}} = - {\mathbf N}_{3{\mathrm b}}$ becomes ${\mathbf N}_{3{\mathrm a}} = {\mathbf N}_{3{\mathrm b}}$ or ${\mathbf N}_1 {\mathbf N}_2 = {\mathbf N}_2 {\mathbf N}_1$.
Thus, such a situation could be incorrectly interpreted as `Abelian’. 
This behavior is, however, simply a result coming from the gauge freedom of the eigenstates and there is no problem in the non-Abelian nature of the non-Abelian band topology.

\section{\label{sec:Considerations}Considerations of using the reference point method in phononic and photonic topological insulators}
\subsection{\label{sec:PathSetting}${\mathbf k}_r - {\mathbf k}_0$ Path and degeneracies}
Although Fig.~\ref{fig:RefPoint} shows the paths between ${\mathbf k}_r$ and ${\mathbf k}_0$ in linear forms, they do not have to be a straight line.
Instead of the straight line [see Fig.~\ref{fig:PathSetting}(a)], the path sometimes should make a detour around degeneracies [see Fig.~\ref{fig:PathSetting}(b)] whichever the degeneracies are zero, one, two, or three dimensions.
Suppose we investigate the topological nature by the three bands indexed as $n$, $n+1$, and $n+2$.
The paths should avoid the degeneracies that at least one of these three bands is concerned with, e.g., the degeneracies by the bands $n-1$ and $n$ or the degeneracies by the bands $n+2$ and $n+3$.
This is because the degeneracies destroy the eigenstates information at ${\mathbf k}_r$.
In Fig.~\ref{fig:PathSetting}(c), Path 1 goes through the pink points (near ${\mathbf k}_0$) while Path 2 avoids them.
The pink points are degeneracies between the second and third bands ($n=0$ and $1$, respectively).
$\left\langle \psi_{{\mathbf k}-\Delta {\mathbf k}}^n \mid \psi_{\mathbf k}^n \right\rangle$ ($n=1,2,3$) at each point on Path 1 are calculated, and only the value with $n=1$ shows the sharp change around the degeneracies [see Fig.~\ref{fig:PathSetting}(d)].
In contrast, their behaviors on Path 2 do not exhibit such a sharp change [see Fig.~\ref{fig:PathSetting}(e)].
In other words, the eigenstates information at ${\mathbf k}_r$ cannot be delivered through the degeneracies because the degeneracies may act as the sink or source of certain topological states, like Weyl points \cite{Lu_2013_NatPhoton,Lu_2015_Science,Park_2020_ACSPhotonics,He_2018_Nature,Peri_20190_NatPhys,He_2020_NatComm,Jia_2019_Science,Yang_2018_Science,Yang_2019_NatComm,Soluyanov_2015_Nature,Fruchart_2018_PNAS,Yang_2017_OptExpress}.

The reference point ${\mathbf k}_r$ also should be fixed such that the overall ${\mathbf k}_r - {\mathbf k}_0$ distances are minimized.
This is because the longer distance of ${\mathbf k}_r - {\mathbf k}_0$ may cause the loss of the Bloch states at ${\mathbf k}_r$ due to the degeneracies or other unknown factors.

\subsection{\label{sec:Freedom_RefPoint}Gauge freedom at the reference point}
Although we examined the topological charges of our topological insulators by applying the reference point method, this method does not determine the exact sign on each topological charge because the sign freedom of the eigenstates at ${\mathbf k}_r$ still remains.
If the eigenstates' signs at ${\mathbf k}_r$ are flipped, some quaternion charges’ signs may be flipped while the other quaternion charges's signs do not change. 
For example, regarding Fig.~\ref{fig:k_i_AfterSignConvention} and \ref{fig:j_AfterSignConvention}, if we flip the signs of ${\mathbf u}_{{\mathbf k}_r}^1$ and ${\mathbf u}_{{\mathbf k}_r}^2$ at ${\mathbf k}_r$, the result in Fig.~\ref{fig:k_i_AfterSignConvention}(e) may change from ${\boldsymbol i}$ to ${-\boldsymbol i}$ while the topological charge of Fig.~\ref{fig:k_i_AfterSignConvention}(c) is still ${\boldsymbol k}$.
Fortunately, in this situation, the signs in Fig.~\ref{fig:j_AfterSignConvention} are also inverted, and therefore the non-Abelian relations between the charges survive. 

\section{\label{sec:Conclusion}Conclusions}
In summary, we have visualized the gauge-dependent property of the non-Abelian charges \cite{Wu_2019_Science,Bouhon_2019_PRB,Bouhon_2020_PRB,Tiwari_2020_PRB,Guo_2021_Nature,Jiang_2021_NatPhys,Peng_2021_arXiv} in phononic and photonic topological insulators.
We have also applied the reference point method, which is equivalent to the basis fixing \cite{Wu_2019_Science,Tiwari_2020_PRB}, to our topological insulators.
Additionally, the consistency of non-Abelian relationship between topological charges with this method was discussed.

Based on the discussions in Section~\ref{sec:Freedom_RefPoint}, we remark the following conclusions:
(i) The gauge fixing by the common reference point does not fix the quaternion charges' signs but fix the relations between the signs.
(ii) The signs of the nodal lines are adjustable by regulating the eigenstates’ signs at the reference point ${\mathbf k}_r$.
(iii) The sign is not an absolute feature of the non-Abelian charge, but it is concerned with the relations between charges.
In other words, the sign relation between different nodal lines is more important than the sign of absolute charges corresponding to each loop.

\begin{acknowledgments}
The work is part-funded by the European Regional Development Fund through the Welsh Government (80762-CU145 (East)).
\end{acknowledgments}

\appendix

\begin{figure*}
    \includegraphics{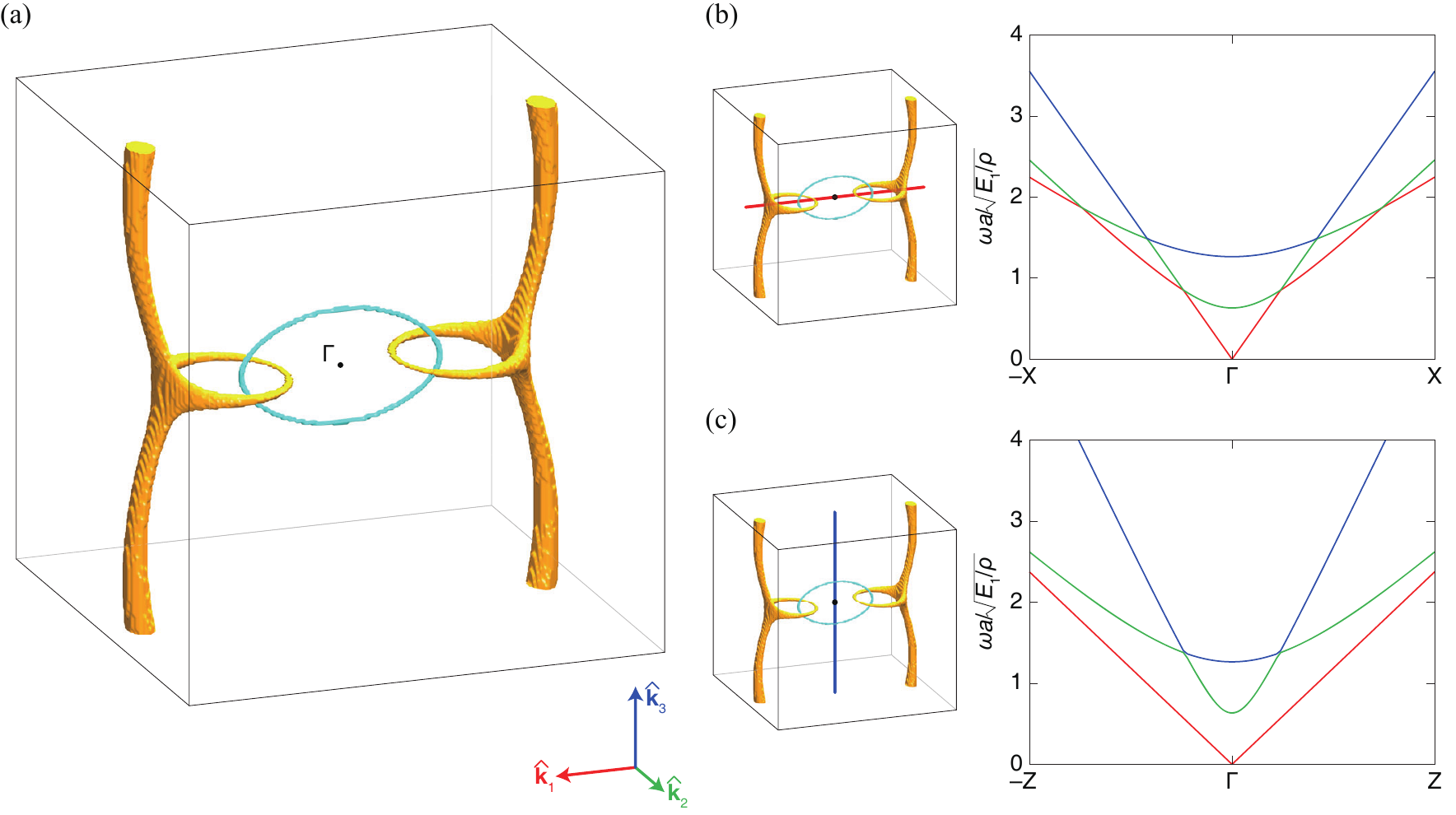}
    \caption{
        \label{fig:NodalLink_Orthotropic}
        Nodal link by the orthotropic elastic material. (a) Set of degeneracies forming a nodal link. The degeneracies by the first and second bands and the second and third bands appear as the orange and cyan nodal lines, respectively. (b),(c) Band structures along ${\mathbf k}_1$ and ${\mathbf k}_3$ directions, respectively.
    }
\end{figure*}
\section{\label{OrthotropicDerivation}Derivation of $3\times3$ Hamiltonian for orthotropic elastic material}

The constitutive equation for an orthotropic elastic material is expressed as ${\boldsymbol \varepsilon} = {\cal C} \cdot {\mathbf s}$, or
\begin{equation}
    \left[
        \begin{array}{ccc}
            \varepsilon_{11} \\ \varepsilon_{22} \\ \varepsilon_{33} \\ 2\varepsilon_{23} \\ 2\varepsilon_{31} \\ 2\varepsilon_{12}
        \end{array}
    \right]
    =
    {\cal C} \cdot
    \left[
        \begin{array}{ccc}
            s_{11} \\ s_{22} \\ s_{33} \\ s_{23} \\ s_{31} \\ s_{12}
        \end{array}
    \right]
    ,
\end{equation}
where ${\boldsymbol \varepsilon} = \left\{ \varepsilon_{ij} \right\}$ and ${\mathbf s} = \left\{ s_{ij} \right\}$ are the Cauchy strain and stress tensors, respectively ($i,j = 1,2,3$).
Each component of the Cauchy strain tensor is expressed as $\varepsilon_{ij} = \left( 1/2 \right) \left( \partial u_i / \partial X_j + \partial u_j / \partial X_i \right)$ where ${\mathbf u} = \left[ u_1 , u_2 , u_3 \right]$ is the displacement vector. $\cal C$ is the compliance tensor:
\begin{equation}
    \cal C
    =
    \left[
        \begin{array}{cccccc}
            \frac 1 {E_1} & -\frac {\nu_{21}} {E_2} & -\frac {\nu_{31}} {E_3} & 0 & 0 & 0 \\
            -\frac {\nu_{12}} {E_1} & \frac 1 {E_2} & -\frac {\nu_{32}} {E_3} & 0 & 0 & 0 \\
            -\frac {\nu_{13}} {E_1} & -\frac {\nu_{23}} {E_2} & \frac 1 {E_3} & 0 & 0 & 0 \\
            0 & 0 & 0 & \frac 1 {G_{23}} & 0 & 0 \\
            0 & 0 & 0 & 0 & \frac 1 {G_{31}} & 0 \\
            0 & 0 & 0 & 0 & 0 & \frac 1 {G_{12}} \\
        \end{array}
    \right]
    ,
\end{equation}
where $E_i$ is the Young's modulus along $i$-direction and $G_{ij} = G_{ji}$ is the shear modulus along $i$-direction on the plane normal to $j$-direction.
$\nu_{ij}$ is the Poisson's ratio, the negative ratio of a transverse strain along $j$-direction to a longitudinal strain along $i$-direction.
Because we assume an orthotropic material, the relation $\nu_{ij} / E_i = \nu_{ji} / E_j$ holds and the compliance tensor $\cal C$ is symmetric.

\begin{table}[h]
    \caption{\label{tab:ParameterData}Parameter-sets to realize the nodal links by orthotropic elastic material.}
    \begin{ruledtabular}
    \begin{tabular}{lc}
        Quantities&Values\\
        \colrule
        Young's moduli (MPa)    &   $E_1 = 50$, $E_2 = 40$, $E_3 = 110$ \\
        Shear moduli (MPa)      &   $G_{12} = 17.5$, $G_{23} = 26.7$\\
                                &   $G_{31} = 28.6$\\
        Poisson's ratios (1)    &   $\nu_{21} = 0.25$, $\nu_{13} = 0.2$, $\nu_{32} = 0.4$ \\
        Density (kg/m\textsuperscript{3})   &   $\rho = 1000$ \\
        Restoring forces (N/m\textsuperscript{3})   &   $f_1 = 0$, $f_2 = 80$, $f_3 = 20$
    \end{tabular}
    \end{ruledtabular}
\end{table}

The wave equation for an elastic material is $- \nabla \cdot {\boldsymbol \sigma} - {\mathbf F}_v = -\rho \ddot{\mathbf u}$, where ${\mathbf F}_v$ is the body force.
Here, we consider the body force as the restoring force expressed as ${\mathbf F}_v = - {\mathbf f} \cdot {\mathbf u}$, i.e.,   ${\mathbf F}_v = - \left[ f_1 u_1 , f_2 u_2 , f_3 u_3 \right]$.
If the displacement vector is expressed as ${\mathbf u} \left( {\mathbf x}, t \right) = {\mathbf u} \left( {\mathbf x} \right) e^{-i \omega t} = {\mathbf u}_{\mathbf k} e^{-i {\mathbf k} \cdot {\mathbf x}} e^{-i \omega t}$, substituting this into the wave equation leads Eq.~(\ref{orthogropic3x3}). In Eq.~(\ref{orthogropic3x3}), $M_{ii}$ and $\lambda_{ij}$ are the components of the stiffness matrix ${\cal S} = {\cal C}^{-1}$ as follows:
\begin{equation}
    \cal S
    =
    \left[
        \begin{array}{cccccc}
            M_{11} & \lambda_{21} & \lambda_{31} & 0 & 0 & 0 \\
            \lambda_{12} & M_{22} & \lambda_{23} & 0 & 0 & 0 \\
            \lambda_{31} & \lambda_{23} & M_{33} & 0 & 0 & 0 \\
            0 & 0 & 0 & G_{23} & 0 & 0 \\
            0 & 0 & 0 & 0 & G_{31} & 0 \\
            0 & 0 & 0 & 0 & 0 & G_{12} \\
        \end{array}
    \right]
    ,
\end{equation}
This orthotropic elastic material forms a nodal link when the parameters in Table \ref{tab:ParameterData} are used.
In the given the momentum space with $-\pi \leq k_1 \leq \pi$, $-\pi \leq k_2 \leq \pi$, and $-\pi \leq k_3 \leq \pi$, we calculated the eigenfrequencies $\omega^n$.
We regarded that two bands $n$ and $n+1$ at point $\mathbf k$ are degenerated if ${\left | \omega^{n+1} - \omega^n \right|} a / \sqrt {E_1 / \rho} < 0.0268$, where $a=1$ m is the constant for normalization.
Sets of the degeneracies formed between the first and second bands and the second and third bands are depicted as orange and cyan nodal lines, respectively, in Fig.~\ref{fig:NodalLink_Orthotropic}(a).
Band structures along ${\mathbf k}_1$ and ${\mathbf k}_3$ directions are also given in Fig.~\ref{fig:NodalLink_Orthotropic}(b),(c), respectively.


\bibliography{SignConvention}

\end{document}